\documentclass[12pt]{article}
\usepackage{amsmath,amsfonts,amssymb}  
\makeatletter
\def\numberbysection{\@addtoreset{equation}{section}
    \def\theequation{\thesection.\arabic{equation}}}
\makeatother

\numberbysection

\newcommand{\be}{\begin{eqnarray}}
\newcommand{\ee}{\end{eqnarray}}
\newcommand{\non}{\nonumber}
\newcommand{\id}{\mathbb{I}}
\newcommand{\tr}{\mathop{\rm tr}\nolimits}
\newcommand{\str}{\mathop{\rm str}\nolimits}
\newcommand{\sgn}{\mathop{\rm sign}\nolimits}
\newcommand{\diag}{\mathop{\rm diag}\nolimits}

\begin{document}

\begin{titlepage}
\strut\hfill UMTG--259
\vspace{.5in}
\begin{center}

\LARGE Open-chain transfer matrices for AdS/CFT\\[1.0in]
\large Rajan Murgan\footnote{Current address: Physics Department, 
Gustavus Adolphus College, Olin Hall, 800 West College Avenue, 
St. Peter, MN 56082 USA}${}^{,}$\footnote{rmurgan@gustavus.edu}
and Rafael I. Nepomechie\footnote{nepomechie@physics.miami.edu}\\[0.8in]
\large Physics Department, P.O. Box 248046, University of Miami\\[0.2in]  
\large Coral Gables, FL 33124 USA\\
      
\end{center}

\vspace{.5in}

\begin{abstract}
We extend Sklyanin's construction of commuting open-chain transfer
matrices to the $SU(2|2)$ bulk and boundary $S$-matrices of AdS/CFT.
Using the graded version of the $S$-matrices leads to a transfer
matrix of particularly simple form.  We also find an $SU(1|1)$
boundary $S$-matrix which has one free boundary parameter.
\end{abstract}

\end{titlepage}

\setcounter{footnote}{0}

\section{Introduction}\label{sec:intro}

The factorizable $SU(2|2)$-invariant bulk $S$-matrix proposed by
Beisert \cite{Be, AFZ} plays a central role in understanding
integrability in the closed string/spin chain sector of AdS/CFT.
Indeed, this $S$-matrix can be used to derive \cite{Be, MM, dL} the
all-loop asymptotic Bethe ansatz equations \cite{BS} and to compute
finite-size effects \cite{BJ}.  (For reviews and further references,
see for example Ref. \cite{reviews}.)

Integrability also extends to the open string/spin chain sector of
AdS/CFT. (See for example \cite{BV}-\cite{MV} and references therein.)
Hofman and Maldacena \cite{HM} have proposed boundary $S$-matrices
corresponding to open strings attached to maximal giant gravitons
\cite{giants} in $AdS_{5}\times S^{5}$.  While there has been some
subsequent work (see for example \cite{CC}-\cite{CY}), the study of
integrability in the open string/spin chain sector is considerably
less-well developed compared with the closed string/spin chain sector.
In particular, corresponding all-loop asymptotic Bethe ansatz
equations have yet to be derived.

An important prerequisite for deriving such Bethe ansatz equations is
to construct a commuting open-chain transfer matrix, which is the main
purpose of this note.  Sklyanin \cite{Sk} long ago made the key
observation that the transfer matrix should be of the ``double-row'' form.
However, because the bulk $S$-matrix is not of the difference form and
has a peculiar crossing property \cite{Ja, AFZ}, it is necessary to
generalize his construction.  Indeed, we argue that the transfer 
matrix contains an unexpected factor (\ref{emm}) which is essential 
for commutativity. This factor can be removed by working instead with 
graded versions of the $S$-matrices.

The $SU(2|2)$ bulk $S$-matrix has an $SU(1|1)$ submatrix which
itself satisfies the Yang-Baxter equation \cite{BS, Be2}.  We find
here a corresponding boundary $S$-matrix which, unlike those found in
\cite{HM}, contains an arbitrary boundary parameter.  The simplicity
of the $SU(1|1)$ bulk and boundary $S$-matrices suggests that they can
serve as useful toy models of the more complicated $SU(2|2)$ case.
 
The outline of this paper is as follows. In Section 
\ref{sec:transfer} we construct two different commuting open-chain transfer 
matrices. The first, constructed with non-graded $S$-matrices, 
contains an unexpected factor; and the second, constructed with 
graded versions of the $S$-matrices, does not have this extra factor. 
In Section \ref{sec:su11} we present the $SU(1|1)$ boundary 
$S$-matrix. We conclude in Section \ref{sec:discuss} with a brief 
discussion of our results. An appendix contains the $SU(2|2)$ bulk
$S$-matrix and explains some of our notation.

\section{Transfer matrix}\label{sec:transfer}

Bulk and boundary $S$-matrices are the two main building blocks of the
transfer matrix.  We assume here that the bulk $S$-matrix is
essentially the one found by Beisert \cite{Be} based on $SU(2|2)$
symmetry, but in a basis \cite{AFZ} where the standard Yang-Baxter
equation (YBE)
\be
S_{12}(p_{1}, p_{2})\, S_{13}(p_{1}, p_{3})\, S_{23}(p_{2}, p_{3})\ =
S_{23}(p_{2}, p_{3})\, S_{13}(p_{1}, p_{3})\, S_{12}(p_{1}, p_{2}) \,.
\label{YBE}
\ee
is satisfied.  We use the standard convention $S_{12} = S \otimes
\id$, $S_{23} = \id \otimes S$, and $S_{13} = {\cal P}_{12}\, S_{23}\,
{\cal P}_{12}$, where ${\cal P}_{12} = {\cal P} \otimes \id$, ${\cal
P}$ is the permutation matrix, and $\id$ is the four-dimensional
identity matrix.  For convenience, this $S$-matrix is given explicitly
in the Appendix. For simplicity, we omit the scalar factor. Hence, 
this matrix has the unitarity property
\be
S_{12}(p_{1}, p_{2})\, S_{21}(p_{2}, p_{1}) = \id \,,
\label{unitarity}
\ee
where $S_{21} = {\cal P}_{12}\, S_{12}\, {\cal P}_{12}$, as well as 
the crossing property \cite{Ja, AFZ}
\be
C_{2}(p_{2})\, S_{12}(p_{1}, \bar p_{2})\, C_{2}(p_{2})^{-1}\, 
S_{12}(p_{1},p_{2})^{t_{2}}  =  \id f(p_{1},p_{2}) \,,
\label{crossing}
\ee 
where $C(p)$ is the matrix 
\be
C(p)= \left( \begin{array}{cccc}
0  & i \sgn(p)  & 0 & 0 \\
- i \sgn(p) &0  &0 & 0 \\
0  &0  &0 & 1 \\
0  &0  &-1 & 0
\end{array} \right) \,,
\label{CCmatrix}
\ee 
and the scalar function $f(p_{1},p_{2})$ is given by 
\be
f(p_{1},p_{2}) = \frac{\left(\frac{1}{x^{+}_{1}} - 
x^{-}_{2}\right)(x^{+}_{1} - x^{+}_{2})}
{\left(\frac{1}{x^{-}_{1}} - 
x^{-}_{2}\right)(x^{-}_{1} - x^{+}_{2})} \,.
\label{ffunc}
\ee
Moreover, $\bar p = -p$ denotes the antiparticle momentum, with 
\be
x^{\pm}(\bar p) = \frac{1}{x^{\pm}(p)} \,.
\label{barp}
\ee
As we shall see, the peculiar dependence of the charge conjugation
matrix $C(p)$ on the sign of $p$ gives rise to a nontrivial factor in
the transfer matrix.

We assume here that the right boundary $S$-matrix $R^{-}(p)$ is
essentially the one found by Hofman and Maldacena \cite{HM} for the
so-called $Y=0$ giant graviton brane, but in a basis \cite{AN} where
the standard (right) boundary Yang-Baxter equation (BYBE) \cite{Ch,
GZ}
\be
S_{12}(p_{1}, p_{2})\, R_{1}^{-}(p_{1})\, S_{21}(p_{2}, -p_{1})\, 
R_{2}^{-}(p_{2}) = 
R_{2}^{-}(p_{2})\, S_{12}(p_{1}, -p_{2})\, R_{1}^{-}(p_{1})\, 
S_{21}(-p_{2}, -p_{1}) 
\label{BYBE}
\ee 
is satisfied. It is a diagonal matrix given by \cite{AN}
\be
R^{-}(p) = \diag \left( e^{-ip}\,, -1 \,, 1 \,, 1 \right) \,.
\label{boundarySright}
\ee
As noted in \cite{HM}, 
\be
x^{\pm}(-p) = - x^{\mp}(p) \,, \qquad \eta(-p) = \eta(p) \,,
\label{negation}
\ee 

From the bulk $S$-matrix, we construct a pair of monodromy matrices
\be
T_{a}(p\,; \{p_{i}\}) &=& S_{a N}(p,p_{N}) \cdots S_{a 1}(p,p_{1}) \,,
\non \\
\widehat T_{a}(p\,; \{p_{i}\}) &=& S_{1 a}(p_{1}, -p) \cdots S_{N 
a}(p_{N}, -p) \,,
\label{monodromy}
\ee
where $\{p_{1}, \ldots, p_{N}\}$ are arbitrary ``inhomogeneities''
associated with each of the $N$ quantum spaces, and the auxiliary
space is denoted by $a$.  (As usual, the quantum-space ``indices'' are
suppressed from the monodromy matrices.)  These matrices obey the
relations
\be
S_{a b}(p_{a}, p_{b})\, T_{a}(p_{a}\,; \{p_{i}\})\,
T_{b}(p_{b}\,; \{p_{i}\})  &=& T_{b}(p_{b}\,; \{p_{i}\})\,
T_{a}(p_{a}\,; \{p_{i}\})\, S_{a b}(p_{a}, p_{b}) \,, \non \\
S_{b a}(-p_{b}, -p_{a})\, \widehat T_{a}(p_{a}\,; \{p_{i}\})\,
\widehat T_{b}(p_{b}\,; \{p_{i}\}) &=& 
\widehat T_{b}(p_{b}\,; \{p_{i}\})\, 
\widehat T_{a}(p_{a}\,; \{p_{i}\})\, S_{b a}(-p_{b}, -p_{a})\,, \non \\
\widehat T_{a}(p_{a}\,; \{p_{i}\})\, S_{b a}(p_{b}, -p_{a})  
T_{b}(p_{b}\,; \{p_{i}\}) &=& 
T_{b}(p_{b}\,; \{p_{i}\})\, S_{b a}(p_{b}, -p_{a})\, 
\widehat T_{a}(p_{a}\,; \{p_{i}\}) 
\label{fundamental}
\ee
as a consequence of the YBE. The ``decorated'' right boundary $S$-matrix given by
\be
{\cal T}^{-}_{a}(p\,; \{p_{i}\}) = T_{a}(p\,; \{p_{i}\})\,
R_{a}^{-}(p)\,  \widehat T_{a}(p\,; \{p_{i}\})
\ee
also satisfies the BYBE, i.e.,
\be
\lefteqn{S_{a b}(p_{a}, p_{b})\, {\cal T}_{a}^{-}(p_{a}\,; \{p_{i}\})\, 
S_{b a}(p_{b}, -p_{a})\, 
{\cal T}_{b}^{-}(p_{b}\,; \{p_{i}\})} \non \\
& &= {\cal T}_{b}^{-}(p_{b}\,; \{p_{i}\})\, S_{a b}(p_{a}, -p_{b})\, {\cal 
T}_{a}^{-}(p_{a}\,; \{p_{i}\})\, 
S_{b a}(-p_{b}, -p_{a}) \,,
\ee 
by virtue of (\ref{BYBE}) and (\ref{fundamental}).

Following Sklyanin \cite{Sk}, we assume that the open-chain transfer 
matrix is of the double-row form
\be
t(p\,; \{p_{i}\}) &=& \tr_{a} R_{a}^{+}(p)\, {\cal T}^{-}_{a}(p\,; 
\{p_{i}\})  \non \\
&=& \tr_{a} R_{a}^{+}(p)\, T_{a}(p\,; \{p_{i}\})\,
R_{a}^{-}(p)\,  \widehat T_{a}(p\,; \{p_{i}\}) \,,
\label{transfer}
\ee 
where the trace is over the auxiliary space, and the left boundary
$S$-matrix $R^{+}(p)$ is chosen to ensure the essential commutativity
property
\be
\left[ t(p\,; \{p_{i}\}) \,, t(p'\,; \{p_{i}\}) \right] = 0 
\label{commutativity}
\ee
for arbitrary values of $p$ and $p'$.  By repeating the (not short)
computation in \cite{Sk}, but now making use of the unitarity and
crossing properties (\ref{unitarity}) and (\ref{crossing}), we find
that the commutativity property is indeed obeyed, provided that
$R^{+}(p)$ satisfies the relation
\be
\lefteqn{S_{2 1}(p_{2}, p_{1})^{{t}_{12}}\, R^{+}_{1}(p_{1})^{{t}_{1}}\, 
C_{1}(-p_{1})\, S_{21}(p_{2}, \overline{-p_{1}})^{t_{2}}\, 
C_{1}(-p_{1})^{-1}\,  R^{+}_{2}(p_{2})^{{t}_{2}}} \non \\
& & =  R^{+}_{2}(p_{2})^{{t}_{2}}\, C_{2}(-p_{2})\, 
S_{12}(p_{1}, \overline{-p_{2}})^{t_{1}}\, C_{2}(-p_{2})^{-1}\,
R^{+}_{1}(p_{1})^{{t}_{1}}\, S_{1 2}(-p_{1}, -p_{2})^{{t}_{12}} \,.
\label{ugly}
\ee
In obtaining this result, we also make use of the identity
\be
f(p_{1},p_{2}) = f(-p_{2}, -p_{1}) 
\ee
which is satisfied by the function defined in (\ref{ffunc}).  The
relation (\ref{ugly}) can be simplified using again the crossing
property (\ref{crossing}). Eventually, we arrive at
\be
\lefteqn{S_{12}(p_{1}, p_{2})\, M_{1}\, R_{1}^{+}(-p_{1})\, S_{21}(p_{2}, -p_{1})\, 
M_{2}\, R_{2}^{+}(-p_{2})} \non \\
& & = M_{2}\, R_{2}^{+}(-p_{2})\,  S_{12}(p_{1}, -p_{2})\, M_{1}\, R_{1}^{+}(-p_{1})\,  
S_{21}(-p_{2}, -p_{1}) \,,
\label{good}
\ee 
where the matrix $M$ is given by
\be
M = C(-p)\, C(p)^{-1} = \diag \left(-1\,, -1\,, 1\,, 1 \right) = M^{-1} \,.
\label{emm}
\ee 
In obtaining this result, we make use of the identities
\be
f(p_{1},p_{2}) = f(\overline{-p_{2}}, \overline{-p_{1}}) 
\ee
and
\be
M_{1}\, S_{12}(p_{1}, p_{2})\, M_{2} = M_{2}\, S_{12}(p_{1}, p_{2})\, 
M_{1} \,.
\ee 

Comparing the $R^{+}(p)$ relation (\ref{good}) with the $R^{-}(p)$ 
relation (\ref{BYBE}), we conclude that the left boundary $S$-matrix 
is given by
\be
R^{+}(p) =  M R^{-}(-p) \,,
\label{boundarySleft}
\ee
where $M$ is given by (\ref{emm}). We emphasize that this matrix $M$,
which arises from the peculiar dependence of the charge conjugation 
matrix on the sign of the momentum, is essential in order for the 
transfer matrix (\ref{transfer}) to have the commutativity property 
(\ref{commutativity}), which we have verified numerically for small 
numbers of sites. A formally similar matrix appears in the construction 
of open-chain transfer matrices for nonsymmetric $R$-matrices \cite{MN}.

The matrix $M$ does not appear if we work instead with corresponding 
graded quantities.\footnote{For the generalization of Sklyanin's 
formalism to graded $S$-matrices, see for example \cite{graded}.}
Indeed, let us make the parity assignments  
\be
p(1) = p(2) = 0\,, \qquad p(3) = p(4) = 1 \,,
\ee
and define the graded bulk $S$-matrix by (see, e.g., \cite{MM})
\be
S^{g}(p_{1}, p_{2}) = {\cal P}^{g}\, {\cal P}\, S(p_{1}, p_{2})\,,
\label{gradedS}
\ee
where ${\cal P}^{g}$ is the graded permutation matrix
\be
{\cal P}^{g} = \sum_{i,j=1}^{4} (-1)^{p(i) p(j)} e_{i\, j} \otimes 
e_{j\, i} \,,
\ee
and $S(p_{1}, p_{2})$ is given in 
the Appendix. We consider the transfer matrix given by
\be
t(p\,; \{p_{i}\}) = \str_{a} R_{a}^{+}(p)\, T_{a}(p\,; \{p_{i}\})\,
R_{a}^{-}(p)\,  \widehat T_{a}(p\,; \{p_{i}\}) \,,
\label{gradedtransfer1}
\ee 
where $\str$ denotes the supertrace, the monodromy matrices are formed
as in (\ref{monodromy}) except with the graded $S$-matrix
(\ref{gradedS}) using the graded tensor product (instead of the
ordinary tensor product), and $R^{-}(p)$ is again given by
(\ref{boundarySright}), which also satisfies the graded BYBE. The
transfer matrix (\ref{gradedtransfer1}) satisfies the commutativity
property (\ref{commutativity}) for $R^{+}(p)$ given by
(\ref{boundarySleft}) with $M=\id$.  That is,
\be
t(p\,; \{p_{i}\}) = \str_{a} R_{a}^{-}(-p)\, T_{a}(p\,; \{p_{i}\})\,
R_{a}^{-}(p)\,  \widehat T_{a}(p\,; \{p_{i}\}) \,.
\label{gradedtransfer2}
\ee 
This transfer matrix evidently has the right structure for formulating the
Bethe-Yang equation on an interval with left and right 
boundaries.\footnote{In the closed string/spin chain sector, it is
necessary to formulate the Bethe-Yang equation using the graded
$S$-matrix in order to properly implement periodic boundary conditions
\cite{MM}.}

\section{$SU(1|1)$ boundary $S$-matrix}\label{sec:su11}

The $SU(2|2)$ bulk $S$-matrix contains an $SU(1|1)$ submatrix which 
itself satisfies the graded YBE, namely, \cite{BS, Be2}
\be
S(p_{1}, p_{2}) = \left( \begin{array}{cccc}
	x^{+}_{1}-x^{-}_{2} &0            &0           &0  \\
	0                 &x^{-}_{1}-x^{-}_{2}    
	&(x^{+}_{1}-x^{-}_{1})\frac{\omega_{2}}{\omega_{1}}  &0  \\
	0                 
	&(x^{+}_{2}-x^{-}_{2})\frac{\omega_{1}}{\omega_{2}}  &x^{+}_{1}-x^{+}_{2}  &0 \\
	0                 &0            &0           &x^{-}_{1}-x^{+}_{2}
\end{array} \right) \,,
\ee
where the parity assignments are $p(1)=0\,, p(2)=1$.
(We are again not concerned here with overall scalar factors.)
Curiously, as already noted by Beisert and Staudacher \cite{BS}, the YBE 
holds even without imposing any constraint between $x^{+}(p)$ and 
$x^{-}(p)$, and without specifying $\omega(p)$.

We find that the corresponding right BYBE has the following diagonal solution
\be
R^{-}(p) = \diag\left(a - x^{+}(p)\,, a + x^{-}(p) \right) \,,
\label{su11sltn}
\ee
where $a$ is an arbitrary boundary parameter.  While the appearance of
boundary parameters is common for boundary $S$-matrices associated
with affine Lie algebras, we emphasize that no such boundary parameter
appears in the $SU(2|2)$ boundary $S$-matrices \cite{HM, AN}.  As is
the case for the bulk, the BYBE is satisfied without imposing any
constraint between $x^{+}(p)$ and $x^{-}(p)$ other than
(\ref{negation}), and without specifying $\omega(p)$ other than
\be
\omega(-p) = \omega(p) \,.
\label{omega}
\ee 

The corresponding commuting open-chain transfer matrix is given by 
(\ref{gradedtransfer1}), where the left boundary $S$-matrix is given by
\be
R^{+}(p) =  R^{-}(-p)\Big\vert_{a \mapsto b} = 
\diag\left(b + x^{-}(p)\,, b - x^{+}(p) \right) \,,
\ee
where $b$ is another arbitrary boundary parameter.  The commutativity
(\ref{commutativity}) holds for arbitrary $x^{+}(p)\,, x^{-}(p)\,,
\omega(p)$ obeying (\ref{negation}), (\ref{omega}).

\section{Discussion}\label{sec:discuss}

We have found that the $SU(2|2)$ bulk and boundary $S$-matrices of
AdS/CFT can be used to construct a commuting open-chain transfer
matrix given by (\ref{transfer}), where $T_{a}(p\,; \{p_{i}\})$ and
$\widehat T_{a}(p\,; \{p_{i}\})$ are given by (\ref{monodromy}), and
$R^{+}(p)$ is given by (\ref{boundarySleft}), which contains the 
unexpected factor $M$ (\ref{emm}).  Alternatively, using graded 
versions of the $S$-matrices, one can construct the simpler transfer 
matrix (\ref{gradedtransfer2}). Moreover, we have found
a new $SU(1|1)$ boundary $S$-matrix (\ref{su11sltn}) which, in
contrast to the $SU(2|2)$ case (\ref{boundarySright}), contains an
arbitrary boundary parameter.

For the $SU(2|2)$ closed chain, a local Hamiltonian can be 
obtained from the closed-chain transfer matrix 
\be
t_{closed}(p\,; \{p_{i}\}) = \tr_{a} T_{a}(p\,; \{p_{i}\})
\ee
by setting all the inhomogeneities equal $p_{i} \equiv p_{0}$, and
taking the logarithmic derivative,
\be
H_{closed} = \frac{d}{dp} \ln t_{closed}(p\,; \{p_{i}=p_{0}\}) \Big\vert_{p=p_{0}} \,.
\ee
As noted by Beisert \cite{Be}, in contrast to the conventional case,
this Hamiltonian depends on the value of $p_{0}$, since the bulk
$S$-matrix does not have the difference property.  Nevertheless, this
Hamiltonian is local, since the $S$-matrix is regular, $S(p_{0},
p_{0}) \propto {\cal P}$, and therefore $t_{closed}(p_{0}\,; 
\{p_{i}=p_{0}\})$ is the one-site shift operator.

It is not clear whether an analogous local Hamiltonian can be 
obtained from the open-chain transfer matrix (\ref{transfer}). Indeed,
in contrast to the conventional homogeneous case \cite{Sk}, $t(p_{0}\,; 
\{p_{i}=p_{0}\})$ is not proportional to the identity. This is due to 
the fact that 
\be
\widehat T_{a}(p_{0}\,; \{p_{i}=p_{0}\}) = 
S_{1 a}(p_{0}, -p_{0}) \cdots S_{N a}(p_{0}, -p_{0}) \,,
\ee
which is not a product of permutation operators, and the fact 
that $R^{-}(p_{0})$ is not proportional to the identity matrix. (This
is true even for the conventional inhomogeneous case.) Hence,
the naive guess
\be
\frac{d}{dp} t(p\,; \{p_{i}=p_{0}\}) \Big\vert_{p=p_{0}}
\label{naive}
\ee
does not give a local Hamiltonian; and multiplying (\ref{naive}) by $t(p_{0}\,; 
\{p_{i}=p_{0}\})^{-1}$ does not help.

It would be interesting to determine the eigenvalues and Bethe ansatz
equations of the $SU(2|2)$ open-chain transfer matrix.  We expect that
the $SU(1|1)$ case will serve as a useful warm-up exercise.

\section*{Acknowledgments}
This work was supported in part by the National Science Foundation
under Grants PHY-0244261 and PHY-0554821.

\begin{appendix}

\section{The $SU(2|2)$-invariant bulk $S$-matrix}

We arrange the bulk $S$-matrix elements into a $16 \times 16$
matrix $S$ as follows,
\be
S(p_{1}, p_{2}) = \sum_{i,i',j,j'=1}^{4} S_{i\, j}^{i' j'}(p_{1}, p_{2})\, 
e_{i\, i'}\otimes e_{j\, j'}\,,
\label{bulkS2}
\ee
where $e_{i j}$ is the usual elementary $4 \times 4$ matrix whose $(i,j)$ 
matrix element is 1, and all others are zero.  Although
(\ref{bulkS2}) is the standard convention, Arutyunov {\it et al.} use
a different convention (see Eq.  (8.4) in \cite{AFZ}), such that our
matrix $S$ is the {\it transpose} of theirs. The nonzero matrix 
elements are \cite{AFZ}
\be
S_{a\, a}^{a\, a}(p_{1}, p_{2}) &=& A\,, \quad 
S_{\alpha\, \alpha}^{\alpha\, \alpha}(p_{1}, p_{2}) = D\,, \non \\
S_{a\, b}^{a\, b}(p_{1}, p_{2}) &=& \frac{1}{2}(A-B)\,, \quad 
S_{a\, b}^{b\, a}(p_{1}, p_{2}) = \frac{1}{2}(A+B) \,, \non \\
S_{\alpha\, \beta}^{\alpha\, \beta}(p_{1}, p_{2}) &=& \frac{1}{2}(D-E)\,, \quad 
S_{\alpha\, \beta}^{\beta\, \alpha}(p_{1}, p_{2}) = \frac{1}{2}(D+E) \,, \non 
\ee
\be 
S_{a\, b}^{\alpha\, \beta}(p_{1}, p_{2}) &=& 
-\frac{1}{2}\epsilon_{a b}\epsilon^{\alpha \beta}\, C \,, \quad
S_{\alpha\, \beta}^{a\, b}(p_{1}, p_{2}) = 
-\frac{1}{2}\epsilon^{a b}\epsilon_{\alpha \beta}\, F \,, \non \\
S_{a\, \alpha}^{a\, \alpha}(p_{1}, p_{2}) &=& G\,, \quad 
S_{a\, \alpha}^{\alpha\, a}(p_{1}, p_{2}) = H \,, \quad 
S_{\alpha\, a}^{a\, \alpha}(p_{1}, p_{2}) = K\,, \quad 
S_{\alpha\, a}^{\alpha\, a}(p_{1}, p_{2}) = L \,,  
\label{bulkS3}
\ee
where $a\,, b \in \{1\,, 2\}$ with $a \ne b$;  
$\alpha\,, \beta \in \{3\,, 4\}$ with $\alpha \ne \beta$; and
\be
A &=& \frac{x^{-}_{2}-x^{+}_{1}}{x^{+}_{2}-x^{-}_{1}}
\frac{\eta_{1}\eta_{2}}{\tilde\eta_{1}\tilde\eta_{2}} \,, \non \\
B &=&-\left[\frac{x^{-}_{2}-x^{+}_{1}}{x^{+}_{2}-x^{-}_{1}}+
2\frac{(x^{-}_{1}-x^{+}_{1})(x^{-}_{2}-x^{+}_{2})(x^{-}_{2}+x^{+}_{1})}
{(x^{-}_{1}-x^{+}_{2})(x^{-}_{1}x^{-}_{2}-x^{+}_{1}x^{+}_{2})}\right]
\frac{\eta_{1}\eta_{2}}{\tilde\eta_{1}\tilde\eta_{2}}\,, \non \\
C &=& \frac{2i x^{-}_{1} x^{-}_{2}(x^{+}_{1}-x^{+}_{2}) \eta_{1} \eta_{2}}
{x^{+}_{1} x^{+}_{2}(x^{-}_{1}-x^{+}_{2})(1 - x^{-}_{1} x^{-}_{2})} 
\,, \qquad
D = -1\,, \non \\
E &=&\left[1-2\frac{(x^{-}_{1}-x^{+}_{1})(x^{-}_{2}-x^{+}_{2})
(x^{-}_{1}+x^{+}_{2})}
{(x^{-}_{1}-x^{+}_{2})(x^{-}_{1} x^{-}_{2}-x^{+}_{1} 
x^{+}_{2})}\right]\,, \non \\
F &=& \frac{2i(x^{-}_{1}-x^{+}_{1})(x^{-}_{2}-x^{+}_{2})(x^{+}_{1}-x^{+}_{2})}
{(x^{-}_{1}-x^{+}_{2})(1-x^{-}_{1} x^{-}_{2})\tilde\eta_{1} \tilde\eta_{2}}\,, 
\non \\
G &=&\frac{(x^{-}_{2}-x^{-}_{1})}{(x^{+}_{2}-x^{-}_{1})}
\frac{\eta_{1}}{\tilde\eta_{1}}\,, \qquad 
H =\frac{(x^{+}_{2}-x^{-}_{2})}{(x^{-}_{1}-x^{+}_{2})}
\frac{\eta_{1}}{\tilde\eta_{2}}\,, \non \\
K &=&\frac{(x^{+}_{1}-x^{-}_{1})}{(x^{-}_{1}-x^{+}_{2})}
\frac{\eta_{2}}{\tilde\eta_{1}}\,, \qquad 
L = \frac{(x^{+}_{1}-x^{+}_{2})}{(x^{-}_{1}-x^{+}_{2})}
\frac{\eta_{2}}{\tilde\eta_{2}}\,, 
\label{bulkS4}
\ee
where 
\be
x^{\pm}_{i} = x^{\pm}(p_{i})\,, \quad 
\eta_{1} = \eta(p_{1}) e^{i p_{2}/2}\,, \quad \eta_{2}=\eta(p_{2})\,, 
\quad \tilde\eta_{1} =\eta(p_{1})\,, \quad \tilde\eta_{2} 
=\eta(p_{2})e^{i p_{1}/2}\,,
\ee
and $\eta(p) = \sqrt{i\left[x^{-}(p)-x^{+}(p)\right]}$. Also,
\be
x^{+}+\frac{1}{x^{+}}-x^{-}-\frac{1}{x^{-}} = \frac{i}{g}\,, \quad 
\frac{x^{+}}{x^{-}} = e^{i p} \,.
\label{eta}
\ee

\end{appendix}


\begin{thebibliography}{99}

\bibitem{Be}
N. Beisert,
``The $su(2|2)$ dynamic $S$-matrix,''
[arXiv:hep-th/0511082];\\
N. Beisert,
``The Analytic Bethe Ansatz for a Chain with Centrally Extended 
$su(2|2)$ Symmetry,''
{\it J. Stat. Mech.}  {\bf 0701}, P017 (2007)
[arXiv:nlin/0610017].

\bibitem{AFZ}
G. Arutyunov, S. Frolov and M. Zamaklar,
`The Zamolodchikov-Faddeev algebra for $AdS_{5} \times S^{5}$ superstring,''
{\it JHEP} {\bf 0704}, 002 (2007)
[arXiv:hep-th/0612229].

\bibitem{MM}
M.J. Martins and C.S. Melo,
``The Bethe ansatz approach for factorizable centrally extended 
$S$-matrices,''
{\it Nucl. Phys. B} {\bf 785}, 246 (2007) 
[arXiv:hep-th/0703086].

\bibitem{dL}
M. de Leeuw,
``Coordinate Bethe Ansatz for the String $S$-Matrix,''
{\it J. Phys. A} {\bf 40}, 14413 (2007)
[arXiv:0705.2369].

\bibitem{BS}
N. Beisert and M. Staudacher,
``Long-range $PSU(2,2|4)$ Bethe ansaetze for gauge theory and strings,''
{\it Nucl. Phys.  B} {\bf 727}, 1 (2005)
[arXiv:hep-th/0504190].

\bibitem{BJ}  Z. Bajnok and R.A. Janik,
``Four-loop perturbative Konishi from strings and finite size effects
for multiparticle states,'' [arXiv:hep-th/0807.0399].

\bibitem{reviews}
A.A. Tseytlin,
``Spinning strings and AdS/CFT duality,''
in Ian Kogan Memorial Volume, {\it From Fields to Strings: Circumnavigating
Theoretical Physics}, M. Shifman, A. Vainshtein, and J. Wheater, eds.
(World Scientific, 2004)
[arXiv:hep-th/0311139];\\ 
N. Beisert,
``The dilatation operator of ${\cal N} = 4$ super Yang-Mills theory and
integrability,''
{\it Phys. Rept.}  {\bf 405}, 1 (2005)
[arXiv:hep-th/0407277];\\
K. Zarembo,
``Semiclassical Bethe ansatz and AdS/CFT,''
{\it Comptes Rendus Physique} {\bf 5}, 1081 (2004)
[{\it Fortsch. Phys.}  {\bf 53}, 647 (2005)]
[arXiv:hep-th/0411191];\\
J. Plefka,
``Spinning strings and integrable spin chains in the AdS/CFT
correspondence,''
{\it Living Rev. Rel.}  {\bf 8}, 9 (2005)
[arXiv:hep-th/0507136];\\
J.A. Minahan, 
``A brief introduction to the Bethe ansatz in ${\cal N}=4$ 
super-Yang-Mills,''
{\it J. Phys.} {\bf A39}, 12657 (2006);\\
K. Okamura,
``Aspects of Integrability in AdS/CFT Duality,''
[arXiv:0803.3999].

\bibitem{BV}
D. Berenstein and S.E. Vazquez,
``Integrable open spin chains from giant gravitons,''
{\it JHEP} {\bf 0506}, 059 (2005)
[arXiv:hep-th/0501078].

\bibitem{MS}
T. McLoughlin and I. Swanson,
``Open string integrability and AdS/CFT,''
{\it  Nucl. Phys. B} {\bf 723}, 132 (2005)
[arXiv:hep-th/0504203].

\bibitem{Ag}
A. Agarwal,
``Open spin chains in super Yang-Mills at higher loops: Some potential
problems with integrability,''
{\it JHEP} {\bf 0608}, 027 (2006)
[arXiv:hep-th/0603067];\\
K. Okamura and K. Yoshida,
``Higher loop Bethe ansatz for open spin-chains in AdS/CFT,''
{\it JHEP} {\bf 0609}, 081 (2006)
[arXiv:hep-th/0604100].

\bibitem{MV}
N. Mann and S.E. Vazquez,
``Classical open string integrability,''
{\it JHEP} {\bf 0704}, 065 (2007)
[arXiv:hep-th/0612038].

\bibitem{HM}
D.M. Hofman and J.M. Maldacena,
``Reflecting magnons,''
{\it JHEP} {\bf 0711}, 063 (2007)
[arXiv:0708.2272].

\bibitem{giants}
J. McGreevy, L. Susskind and N. Toumbas,
``Invasion of the giant gravitons from anti-de Sitter space,''
{\it JHEP} {\bf 0006}, 008 (2000)
[arXiv:hep-th/0003075];\\
M.T. Grisaru, R.C. Myers and O. Tafjord,
``SUSY and Goliath,''
{\it JHEP} {\bf 0008}, 040 (2000)
[arXiv:hep-th/0008015];\\
A. Hashimoto, S. Hirano and N. Itzhaki,
``Large branes in AdS and their field theory dual,''
{\it JHEP} {\bf 0008}, 051 (2000)
[arXiv:hep-th/0008016].

\bibitem{CC}
H.Y. Chen and D.H. Correa,
``Comments on the Boundary Scattering Phase,''
{\it JHEP} {\bf 0802}, 028 (2008)
[arXiv:0712.1361].

\bibitem{ABR}
C. Ahn, D. Bak and S.J. Rey,
``Reflecting Magnon Bound States,''
{\it JHEP} {\bf 0804}, 050 (2008)
[arXiv:0712.4144].

\bibitem{AN}
C. Ahn and R.I. Nepomechie,
``The Zamolodchikov-Faddeev algebra for open strings attached to giant
gravitons,''
{\it JHEP} {\bf 0805}, 059 (2008)
[arXiv:0804.4036].

\bibitem{MuN}
R. Murgan and R.I. Nepomechie,
``$q$-deformed $su(2|2)$ boundary $S$-matrices via the ZF algebra,''
{\it JHEP} {\bf 0806}, 096 (2008)
[arXiv:0805.3142].
  
\bibitem{BL}
N. Beisert and F. Loebbert,
``Open Perturbatively Long-Range Integrable $gl(N)$ Spin Chains,''
[arXiv:0805.3260]. 

\bibitem{Pa}
L. Palla,
``Issues on magnon reflection,''
[arXiv:0807.3646].

\bibitem{CY}
D.H. Correa and C.A.S. Young,
``Reflecting magnons from D7 and D5 branes,''
[arXiv:0808.0452].

\bibitem{Sk}
E.K. Sklyanin, 
``Boundary conditions for integrable quantum systems,''
{\it J. Phys.} {\bf A21}, 2375 (1988).

\bibitem{Ja}
R.A. Janik,
``The $AdS_{5} \times S^{5}$ superstring worldsheet $S$-matrix and crossing 
symmetry,''
{\it Phys. Rev.} {\bf D73}, 086006 (2006)
[arXiv:hep-th/0603038].

\bibitem{Be2}
N. Beisert,
``An $SU(1|1)$-Invariant $S$-Matrix with Dynamic Representations,''
{\it Bulg. J. Phys.} {\bf 33S1} (2006) 371
[arXiv:hep-th/0511013].

\bibitem{Ch} 
I.V. Cherednik, 
``Factorizing particles on a half line and root systems,''
{\it Theor. Math. Phys.} {\bf 61}, 977 (1984).

\bibitem{GZ}
S. Ghoshal and A.B. Zamolodchikov, 
``Boundary $S$-Matrix and Boundary State in Two-Dimensional 
Integrable Quantum Field Theory,''
{\it Int. J. Mod. Phys.} {\bf A9}, 3841 (1994) 
[arXiv:hep-th/9306002].

\bibitem{MN}
L. Mezincescu and R.I. Nepomechie, 
``Integrable open spin chains with nonsymmetric $R$ matrices,'' 
{\it J. Phys.} {\bf A24}, L17 (1991). 

\bibitem{graded}
A. Foerster and M. Karowski,
``The supersymmetric t-J model with quantum group invariance,''
{\it Nucl. Phys.} {\bf B408}, 512 (1993);\\
A. Gonz\'alez-Ruiz,
``Integrable open-boundary conditions for the supersymmetric t-J 
model. The quantum group invariant case,''
{\it Nucl. Phys.} {\bf B424}, 468 (1994) 
[arXiv:hep-th/9401118];\\
R.H. Yue, H. Fan and B.Y. Hou,
``Exact diagonalization of the quantum supersymmetric $SU_q(n|m)$ 
model,''
{\it Nucl. Phys.} {\bf B462}, 167 (1996)
[{\tt cond-mat/9603022}];\\
M. Shiroishi and M. Wadati,
``Integrable Boundary Conditions for the One-Dimensional Hubbard 
Model,''
{\it J. Phys. Soc. Jpn.} {\bf 66}, 2288 (1997)
[arXiv:cond-mat/9708011];\\
A.J. Bracken, X.-Y. Ge, Y.-Z. Zhang and H.-Q. Zhou,
``Integrable open-boundary conditions for the $q$-deformed 
supersymmetric $U$ model of strongly correlated electrons,''
{\it Nucl. Phys.} {\bf B516}, 588 (1998) 
[arXiv:cond-mat/9710141];\\
X.-W. Guan,
``Algebraic Bethe ansatz for the one-dimensional Hubbard model 
with open boundaries,''
{\it J. Phys. A} {\bf 33}, 5391 (2000)
[arXiv:cond-mat/9908054];\\
D. Arnaudon, J. Avan, N. Cramp\'e, A. Doikou, L. Frappat and E. 
Ragoucy,
``General boundary conditions for the sl(N) and $sl(M|N)$ open spin 
chains,''
{\it J. Stat. Mech.} {\bf P08005}, 1 (2004)
[{\tt math-ph/0406021}].




\end{thebibliography}
\end{document}